\documentclass[titlepage]{article}

\usepackage{newunicodechar}
\newunicodechar{∗}{*}
\usepackage{textgreek}
\usepackage{mhchem}
\usepackage{xspace}
\usepackage[inline, shortlabels]{enumitem}
\usepackage{graphicx}
\usepackage{bm}
\usepackage{titling}
\usepackage[noblocks,affil-it]{authblk}

\usepackage{lineno}
\usepackage{siunitx}
\usepackage{natbib}
\usepackage[caption=false]{subfig}
\newcommand{\phantomsubfloat}[1]{
    {\captionsetup[subfigure]{labelformat=empty}
        \subfloat[][]{#1}
    }}
\newcommand{\caplabel}{(\alph*)}
\newcommand{\capref}{(\alph*)}
\usepackage{glossaries-extra}
\usepackage[colorlinks,allcolors=blue]{hyperref}
\usepackage{cleveref}

\crefname{figure}{Fig.}{Figs.}
\Crefname{figure}{Fig.}{Figs.}
\crefname{equation}{Eq.}{Eqs.}
\Crefname{equation}{Eq.}{Eqs.}

\bibstyle{nature}
\newcommand*\nestedglsentry[1]{\protect\ifglsused{#1}{\glsentryshort{#1}}{\glsentrylong{#1}}}

\renewcommand{\citet}[1]{Ref.~\citenum{#1}}

\newcommand{\oned}{$1$D\xspace}
\newcommand{\supp}{Supplementary material}
\newcommand{\supcref}[1]{\supp, \cref{#1}\xspace}
\newcommand{\PP}{$P_\mathrm{in}P_\mathrm{out}$\xspace}
\newcommand{\PS}{$P_\mathrm{in}S_\mathrm{out}$\xspace}
\newcommand{\SP}{$S_\mathrm{in}P_\mathrm{out}$\xspace}
\renewcommand{\SS}{$S_\mathrm{in}S_\mathrm{out}$\xspace}

\newcommand{\mathPS}{P_\mathrm{in}S_\mathrm{out}}

\newcommand{\mathSS}{S_\mathrm{in}S_\mathrm{out}}
\newcommand{\higgs}{Higgs\xspace}

\newcommand{\Higgs}{Higgs\xspace}
\newcommand{\goldstone}{Goldstone\xspace}

\newcommand{\params}{\theta}

\newcommand{\Tc}{70}

\newabbreviation{shg}{SHG}{second harmonic generation}
\newabbreviation{isrs}{ISRS}{impulsive stimulated raman scattering}
\newabbreviation{trshg}{tr-SHG}{time-resolved \nestedglsentry{shg}}
\newabbreviation{rashg}{RA-SHG}{rotational anisotropy \nestedglsentry{shg}}
\newabbreviation{dft}{DFT}{density functional theory}
\newabbreviation{ins}{INS}{inelastic neutron scattering}
\newabbreviation{lswt}{LSWT}{linear spin wave theory}
\newabbreviation{qcp}{QCP}{quantum critical point}
\newabbreviation{lm}{LM}{Levenberg-Marquardt}
\newabbreviation{opa}{OPA}{optical parametric amplifier}

\newcommand\fakesection[1]{
  \refstepcounter{section}
  \addcontentsline{toc}{section}{\protect\numberline{\thesection}#1}
  \sectionmark{#1}
}
 
\begin{document}

\title{\Higgs-mode electromagnon in the spin-spiral multiferroic \ce{CuBr2}}
\author[1*$\dag$]{Bryan T. Fichera}
\author[1*]{Ajesh Kumar}
\author[1,2*]{Baiqing Lv}
\author[1]{Zongqi Shen}
\author[1$\ddag$]{Karna Morey}
\author[1]{Qian Song}
\author[1$\|$]{Batyr Ilyas}
\author[1]{Tianchuang Luo}
\author[1]{Riccardo Comin}
\author[1]{T. Senthil}
\author[1$\P$]{Nuh Gedik}
\affil[1]{Department of Physics, Massachusetts Institute of Technology, Cambridge, Massachusetts 02139, United States}
\affil[2]{Department of Physics and Astronomy, Shanghai Jiao Tong University, Shanghai, 200240, China}
\affil[*]{These authors contributed equally to this work}
\affil[$\dag$]{Present address: Materials Science Division, Argonne National Laboratory, Lemont, IL 60439, United States}
\affil[$\ddag$]{Present address: Department of Physics, Stanford University, Stanford, CA 94305, United States}
\affil[$\|$]{Present address: Department of Physics, University of California at Berkeley and Material Science Division, Lawrence Berkeley National Laboratory, Berkeley, California 94720, United States}
\affil[$\P$]{gedik@mit.edu}

\maketitle
 
\begin{abstract}
Below a continuous symmetry breaking phase transition, the relevant collective excitations are due to longitudinal and transverse fluctuations of the order parameter, which are referred to as \higgs and \goldstone modes, respectively.
In solids, these modes may take on a different character than the equivalent excitations in particle physics due to the diverse vacuum states accessible in condensed matter.
However, the \higgs mode in particular is quite difficult to observe experimentally as it decays quickly into the lower-energy \goldstone bosons and thus has a negligible lifetime in most systems.
In this work, we report evidence for a novel \higgs mode in the multiferroic material \ce{CuBr2}, which shows up as a coherent oscillation in the \glsfmtlong{trshg} signal upon excitation with a femtosecond light pulse.
Since the spiral spin order in \ce{CuBr2} induces a nonzero electric dipole moment in equilibrium, the \higgs mode---which is due to fluctuations in the amplitude of the on-site spin expectation value---is an electromagnon, and thus acquires an inversion quantum number of \num{-1}. 
This is in stark contrast to the \higgs boson of particle physics, which has even parity.
Moreover, the excitation described here represents an entirely new type of electromagnon, distinct from the traditional electromagnon in \glsfmtlong{lswt} which is due to the \goldstone mode.
 \end{abstract}

\section{Introduction}
\begin{figure}
\centering
\phantomsubfloat{\label{fig:fig1a}}
\phantomsubfloat{\label{fig:fig1b}}
\phantomsubfloat{\label{fig:fig1c}}
\makebox[\linewidth]{\includegraphics[width=89mm]{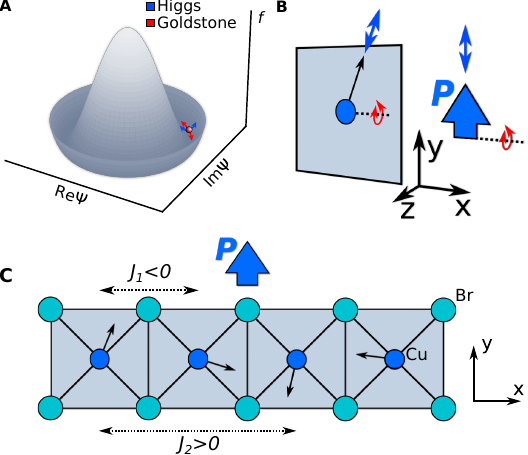}}
\captionsetup{singlelinecheck=off}
\caption[]{
\label{fig:fig1}
\begin{enumerate*}[label=\caplabel, ref=\capref]
\item Mexican hat potential with \higgs and \goldstone modes indicated.
\item $q=0$ electromagnons in the quasi-\oned spin-spiral in the spin (left) and charge (right) sectors.
The \higgs and \goldstone modes are shown in blue and red, respectively.
A second \goldstone mode, corresponding to uniform rotations of the spins about the $z$ axis (which does not affect the polarization $\vec{P}$), is not shown.
\item Magnetic ground state of \ce{CuBr2}.
The macroscopic polarization due to the spin order is depicted with a blue arrow.
The axis labelled $x$ is parallel to the nominal $b$ axis of the crystal structure.
\end{enumerate*}
}
\end{figure}

When the ground state of a given theory fails to respect one of its symmetries, that symmetry is said to have been broken spontaneously\citep{durr_zur_1959,nambu_axial_1960}.
The low-energy excitations of this ground state may then be described as excitations of the order parameter either within the subspace of degenerate ground states or perpendicular to it; these excitations are referred to as \goldstone and \higgs modes, respectively\citep{pekker_amplitudehiggs_2015} (see \cref{fig:fig1a}).
This paradigm describes many fundamental phenomena in both particle physics and condensed matter, and the study of these modes has thus emerged as an essential pursuit in both contexts.

A rich interplay exists between these two fields due to the fact that in particle physics we are limited to a single theory (the standard model), but in condensed matter, the theory is determined by the particular system of interest and may differ significantly from one material to another.
Thus, various exotic species of \higgs modes may be studied simply by exploring different material systems with spontaneous symmetry breaking.
An important example is in multiferroics, where it has been predicted\citep{matsumoto_electromagnon_2014,matsumoto_electromagnon_2015} that the \higgs mode of the magnetic order (corresponding to modulations in the amplitude of the on-site spin expectation value, see \cref{fig:fig1b}) should couple to the macroscopic polarization as an electromagnon, and thus acquire a negative parity eigenvalue.
This is not the case for the Higgs boson of the standard model, which is of even parity\citep{atlas_collaboration_determination_2015}.
In addition to its connection to particle physics, the excitation described here is also fundamentally different from the traditional electromagnon in multiferroics (which is due to the (pseudo-)\goldstone mode\citep{katsura_dynamical_2007}), and is thus of great interest for magnetoelectric device applications.
Unfortunately, like in particle physics, the \higgs mode in condensed matter is difficult to observe since it may quickly decay into \goldstone bosons upon excitation\citep{jain_higgs_2017}, and the existence of this mode in real multiferroic systems has thus remained an important open question.

In this work, we report evidence for this mode in \ce{CuBr2}, a quasi-\oned, spin-spiral multiferroic (see \cref{fig:fig1c}), observed by launching a coherent oscillation of this mode with a near-infrared light pulse and measuring the induced modulations in the electric polarization using a delayed \gls{shg} probe pulse (\cref{fig:fig2a}), similar to \citet{gao_giant_2024}.
We find, as expected, that the mode modulates the macroscopic polarization along the static ordering direction, and that the frequency of the mode decreases on approaching the critical temperature of the multiferroic order.
These results provide conclusive evidence for the existence of this electromagnon in \ce{CuBr2}, solving a longstanding question\citep{matsumoto_electromagnon_2014} and paving the way for future study of the \higgs mode in novel condensed-matter contexts.
 
\section{Results}
\subsection{Equilibrium}
The low-energy spin Hamiltonian of \ce{CuBr2} is well approximated by the so-called frustrated \oned XXZ spin chain, where localized spin-$1/2$ electrons interact ferromagnetically ($J_1 < 0$) with nearest neighbors but antiferromagnetically ($J_2 > 0$) with next-nearest neighbors (see \cref{fig:fig1c}).
When these interaction strengths are of comparable magnitude, the ground state is an incommensurate magnetic spiral, where the ordering wavevector is directed along the chain direction and has the appropriate magnitude so as to balance the two competing interaction terms.
According to theory developed by Katsura, Nagaosa, and Balatsky\citep{katsura_spin_2005}, in the presence of finite spin-orbit coupling this ground state induces an electric polarization at each site $n$ given by
\begin{equation}\label{eq:scrosss}
\vec{P}^n \propto \hat{x}\times(\vec{S}^{n} \times \vec{S}^{n+1}),
\end{equation}
where we have set the chain direction to lie along $\hat{x}$.
If the spins lie in the $xy$ plane, then \cref{eq:scrosss} induces a macroscopic electric polarization which is equal for each bond and directed purely along the $\hat{y}$ direction (\cref{fig:fig1c}).

According to powder neutron diffraction, this spiral magnetic phase is realized in \ce{CuBr2} below $T_c\approx\qty{\Tc}{K}$ \citep{zhao_cubr2_2012,wang_nmr_2018,zhao_pressure_2019,
zhang_giant_2020}, with the propogation vector in reciprocal lattice units given by $\vec{k} = (0, k_y, 0.5)$, with $k_y\sim 0.235$\citep{zhao_cubr2_2012,lee_investigation_2012}.
A pyroelectric current turns on at this temperature as well, indicating a macroscopic electric polarization density $|\vec{P}_0|$ of about \qty{8}{\mu C / m^2} at \qty{10}{K}\citep{zhao_cubr2_2012}.

\begin{figure}
\centering
\phantomsubfloat{\label{fig:fig2a}}
\phantomsubfloat{\label{fig:fig2b}}
\phantomsubfloat{\label{fig:fig2c}}
\includegraphics[width=89mm]{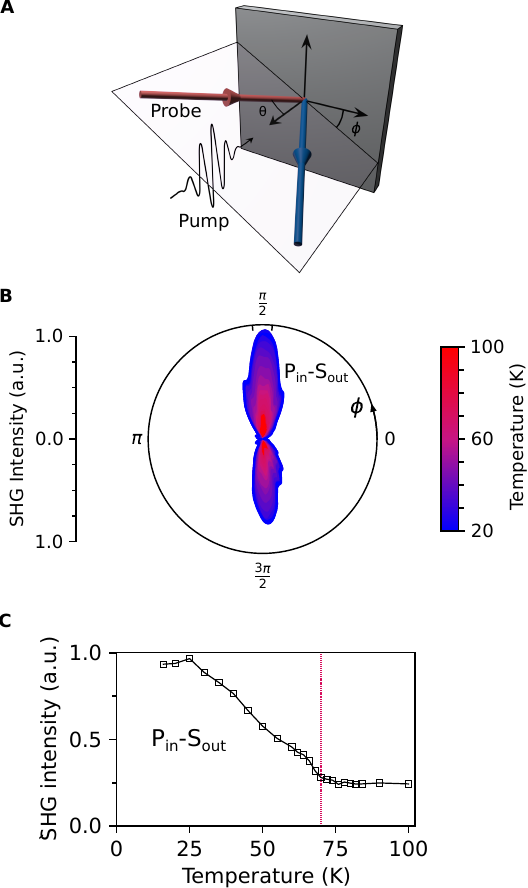}
\captionsetup{singlelinecheck=off}
\caption[]{
\label{fig:fig2}
\begin{enumerate*}[label=\caplabel, ref=\capref]
\item Schematic of the \glsfmtshort{trshg} experimental geometry.
The sample normal is perpendicular to the $(001)$ lattice plane, with the $\phi=0$ axis coincident with the chain direction in \cref{fig:fig1c}.
\item \Gls{shg} intensity as a function of temperature in the \PS polarization channel.
\item Integrated \gls{shg} intensity in the region near $\pi/2$ of \ref{fig:fig2b} marked by the tick marks.
The red dashed line indicates the multiferroic critical temperature $T_c\approx\qty{\Tc}{K}$.
\end{enumerate*}
}
\end{figure}

In a generalized Ginzburg-Landau theory, the \gls{shg} susceptibility tensor $\chi_{ijk}$ is linearly proportional to this polarization:
\begin{equation}\label{eq:glshg}
\chi_{ijk}(T<T_c) = \chi_{ijkl}(T>T_c)P_{0l} = \chi_{ijky}(T>T_c)P_0,
\end{equation}
where $\chi_{ijkl}$ is some unknown tensor with the symmetry of the high temperature phase\citep{sa_generalized_2000}, and we have used that $\vec{P}_0 || \hat{y}$.
\Cref{fig:fig2} shows the temperature dependence of the \gls{shg} intensity in \ce{CuBr2}, indicating a pronounced, order parameter-like enhancement of the \gls{shg} intensity at $T_c$ due to \cref{eq:glshg}.
Note that other contributions to the \gls{shg} intensity due to e.g. magnetic dipole, surface electric dipole, and electric quadrupole terms are allowed above and below $T_c$ and thus cannot explain the pronounced intensity increase below $T_c$.
In addition, the $c$-type electric dipole term purely due to the magnetic order\citep{birss} is also not allowed by the magnetic point group of the incommensurate spin spiral (see \supcref{sup:magshg}).
 
\subsection{Nonequilibrium}
\begin{figure}
\centering
\phantomsubfloat{\label{fig:fig3a}}
\phantomsubfloat{\label{fig:fig3b}}
\phantomsubfloat{\label{fig:fig3c}}
\phantomsubfloat{\label{fig:fig3d}}
\makebox[\linewidth]{\includegraphics[width=183mm]{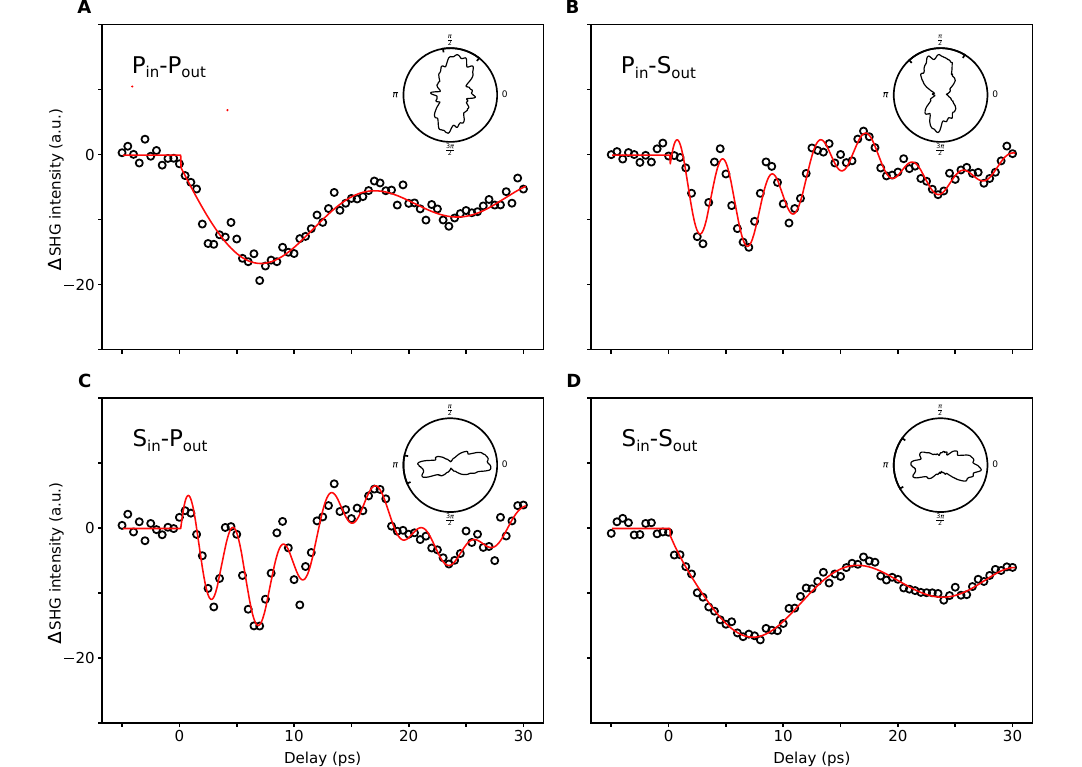}}
\captionsetup{singlelinecheck=off}
\caption[]{
\label{fig:fig3}
\begin{enumerate*}[label=\caplabel, ref=\capref]
\item[] Pump-induced change in the \gls{shg} intensity at \qty{15}{K} in the four polarization channels \item \PP \item \PS \item \SP, and \item \SS.
Insets depict the static \gls{shg} intensity in each polarization channel.
The time-domain signals are computed by performing an azimuthal integration at each delay of the full \gls{shg} pattern over the angles specified by the additional tick marks in each inset.
\end{enumerate*}
}
\end{figure}

Having thus demonstrated that the \gls{shg} intensity is a direct probe of the electric polarization in \ce{CuBr2}, we proceed to investigate the low-energy collective excitations in the multiferroic phase of this material.
In general, the \higgs-mode electromagnon considered in this work may be identified in \gls{trshg} via two defining characteristics.
First of all, since the \higgs mode is directly related to the multiferroic transition at $T_c$, the energy of this mode should soften as the temperature of the sample approaches this value.
Second, the \higgs mode is due to uniform fluctuations in the overall amplitude of the onsite spin expectation value\citep{matsumoto_electromagnon_2014}, which, according to \cref{eq:scrosss}, only affects the longitudinal component of the ferroelectric polarization.
According to \cref{eq:glshg}, the \higgs mode should therefore modulate all of the elements of $\chi_{ijk}$ equally, and should thus appear in all four \gls{shg} polarization channels (\PP, \PS, \SP, and \SS, each of which measures a different linear combination of $\chi_{ijk}$ elements\citep{boyd}).
Here, $P$- ($S$-)polarization refers to electric field oscillation parallel (perpendicular) to the plane of incidence.
In contrast, transverse electromagnons due to pseudo-\goldstone modes \emph{rotate} the polarization vector relative to equilibrium\citep{katsura_dynamical_2007} and thus (per \cref{eq:glshg}) affect different elements of $\chi_{ijk}$ differently.

Keeping these features in mind, we proceed to excite coherent oscillations in \ce{CuBr2} with a \qty{150}{fs} near-infrared pump pulse, and then probe the \gls{shg} intensity with a second pulse delayed in time by an amount $\Delta t$ (see \supcref{sup:excitationmechanism}).
The results are shown in \cref{fig:fig3}.
Two oscillations, with different dependencies on the \gls{shg} polarization channel, may be observed.
One high-frequency mode ($\nu \sim \qty{0.23}{THz}$, $\hbar \omega \sim \qty{1.0}{meV}$), which is only observed in the crossed polarization channels \PS and \SP, may, in light of the above discussion, be assigned to a transverse electromagnon.
A second, low-frequency mode ($\nu \sim \qty{0.05}{THz}$, $\hbar \omega \sim \qty{0.20}{meV}$), which occurs in all polarization channels, may instead be assigned to the longitudinal electromagnon considered in \citet{matsumoto_electromagnon_2014}.
Both of these energies are too low to be observed with typical \si{THz} or neutron spectroscopies, yet they are readily apparent in the \gls{trshg} data due to the pump-probe nature of the experiment.

\begin{figure}
\centering
\phantomsubfloat{\label{fig:fig4a}}
\phantomsubfloat{\label{fig:fig4b}}
\makebox[\linewidth]{\includegraphics[width=89mm]{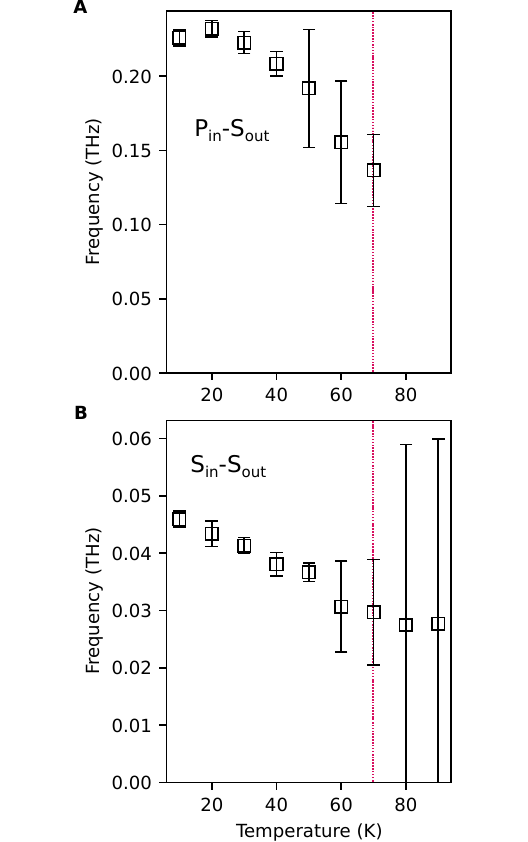}}
\captionsetup{singlelinecheck=off}
\caption[]{
\label{fig:fig4}
\begin{enumerate*}[label=\caplabel, ref=\capref]
\item[] Temperature dependece of the frequencies extracted from the \item \PS and \item \SS time-domain signals (\supcref{sup:timedomain}) in a damped harmonic oscillator model.
Error bars denote \qty{95}{\percent} confidence intervals estimated within a parametric bootstrap (see \supcref{sup:errorbars}).
Red dashed lines indicate the multiferroic critical temperature $T_c\approx\qty{\Tc}{K}$.
\end{enumerate*}
}
\end{figure}

To confirm that these modes are directly involved in the multiferroic phase transition, we measure the pump-induced change in the \gls{shg} intensity as a function of $\Delta t$ for a series of temperatures approaching $T_c$ (see \supcref{fig:timedomain}).
By fitting the respective time-domain traces to damped harmonic oscillators (see \supcref{sup:timedomain}), we can extract the natural frequency of each collective mode as a function of temperature (see \cref{fig:fig4}).
Both modes (\cref{fig:fig4a,fig:fig4b}) exhibit a pronounced softening on approaching $T_c$, confirming their direct involvement in the multiferroic transition.
We also note that both modes disappear above $T_c$, which is sensible given that the macroscopic polarization $\vec{P}_0$ also disappears above this temperature (see \supcref{sup:nooscillationabovetc}).

To further clarify the microscopic origin of these polarization oscillations, we perform \gls{dft}+U and finite-displacement lattice dynamics calculations (see \supcref{sup:phonons}) to compare their energies with those of the zone-center phonon modes.
The lowest zone-center optical phonon in this calculation appears at \qty{7.4}{meV}, in excellent agreement with Raman spectroscopy\citep{wang_observation_2017}, and the calculated acoustic phonon branches (which agree with \gls{ins}\citep{wang_observation_2017}) disperse too rapidly to form a zone-folded acoustic phonon mode at the $\Gamma$ point with an energy low enough to match the frequencies observed in the \gls{trshg} experiment.
Furthermore, acoustic phonon modes close to the $\Gamma$ point could lead to coherent oscillations whose wavevector is obtained by momentum matching with the probe.
However, the calculated mode frequency differs by two orders of magnitude from the measured value (see \supcref{sup:gamma_phonons}), and neither of the modes observed in the \gls{shg} measurement are present in time-resolved reflectivity (see \supcref{fig:pumpprobe}), which would be expected.
Thus, the modes observed in \cref{fig:fig3,fig:fig4} are not phonons.
The only remaining possibility (see \supcref{sup:excluded}) is that these modes are magnons of the incommensurate spin spiral, which imprint themselves on the polarization via \cref{eq:scrosss}; i.e., they are electromagnons.
 
\section{Discussion}
Let us consider the origin of these two electromagnons in \ce{CuBr2} from the perspective of \gls{lswt}.
To begin, note that uniform rotations of the spin spiral about the $z$-axis (corresponding to gapless phasons at zero momentum) do not affect the polarization according to \cref{eq:scrosss}.
There is thus only one spin boson in \gls{lswt} which couples to the polarization in the spiral magnetic phase of \ce{CuBr2} (see \supcref{sup:pimpliess}); it is the so-called pseudo-\goldstone mode of the magnetic order\citep{katsura_dynamical_2007}, which corresponds to a rotation of the spin plane about the chain direction (\cref{fig:fig1b}).
The resulting electromagnon (which, according to \cref{fig:fig1b}, corresponds to a change in polarization $\delta\vec{P}$ along $z$) has zero energy if the system is isotropic about the chain axis, but in the presence of an anisotropy term it acquires a finite energy, which is expected to be in the few-meV range in \ce{CuBr2} (see \supcref{sup:anisotropyenergy}).
This is comparable to the observed value for our high-frequency oscillation (\qty{1.0}{meV}), confirming our previous identification of this mode as a transverse electromagnon.

The observation of a \emph{second} electromagnon in the \gls{trshg}, however, is impossible to explain in \gls{lswt}, and represents the most striking aspect of this work.
We previously argued that this oscillation is due to a modulation in the longitudinal component of the electric polarization (i.e. $\delta\vec{P} || \hat{y}$, see \cref{fig:fig1b}), since it is the \emph{overall} \gls{shg} intensity which is modulated in \cref{fig:fig3}, irrespective of the polarization channel.
We may also argue the same fact by observing that there are only three normal modes of the electric polarization which occur at the $\Gamma$ point in the Brillouin zone: two transverse modes with $\delta\vec{P}$ along $\hat{x}$ and $\hat{y}$ in \cref{fig:fig1b}, and one longitudinal mode with $\delta\vec{P}$ along $\hat{z}$.
Since the $\delta\vec{P} || \hat{z}$ transverse mode is already accounted for by the pseudo-\goldstone mode, that leaves only $\delta\vec{P} || \hat{x}$ and $\delta\vec{P} || \hat{y}$ as possibilities.
The $\delta\vec{P} || \hat{x}$ mode does not couple to the spin order in this compound\citep{katsura_dynamical_2007}, and in any case is not observable in the geometry of our experiment (see \supcref{sup:nopxmode}).
Thus, the only polarization oscillation which is consistent with the observation of a second mode is a longitudinal oscillation with $\delta\vec{P} || \hat{y}$.

Naively, electromagnons with $\delta\vec{P} || \hat{y}$ do not exist in \gls{lswt}.
The key insight, however, is that \gls{lswt} neglects dynamics associated with the magnitude of the onsite spin expectation value.
By \cref{eq:scrosss}, such dynamics change the magnitude of the induced polarization only, not its direction; i.e., they induce longitudinal oscillations with $\delta\vec{P} || \hat{y}$.
In fact, it is possible to show (see \supcref{sup:pimpliess}) that longitudinal oscillations of the induced polarization \emph{necessarily} involve modulations in the amplitude of the onsite spin excpectation value; that is, the only mode which couples to $\delta\vec{P} || \hat{y}$ is the \higgs mode of the magnetic spiral.
Naturally, this mode should soften on approaching $T_c$, in agreement with \cref{fig:fig4} and \cref{fig:timedomain}.
The \qty{0.05}{THz} oscillation observed in our experiment, which is of similar energy to the \higgs mode in non-ferroelectric quantum magnets\citep{ruegg_quantum_2008,hong_higgs_2017}, is therefore direct evidence for this mode in \ce{CuBr2}, with the additional information that it couples to the electric polarization (i.e., it is an electromagnon) via \cref{eq:scrosss}.

Let us make two important remarks about this mode in \ce{CuBr2}.
First, we note that this mode is fundamentally distinct from the \higgs mode in non-multiferroic magnets---since the multiferroic order breaks inversion symmetry, the \higgs mode of this phase has a parity quantum number $l=-1$ rather than $+1$.
Second, we remark that the \higgs mode in \ce{CuBr2} can in principle decay---as in non-multiferroic magnets---quite rapidly into the \goldstone modes of the magnetic order, which in the spin-spiral phase of \cref{fig:fig1c} are gapless and correspond to uniform rotations of each spin about the $\hat{z}$ direction.
The \higgs mode should therefore not exist as a well-defined quasiparticle unless these decay channels are quenched by some mechanism.

In non-multiferroic magnets, two such mechanisms have been identified.
First, the \higgs mode may be stabilized by bringing the system close to a \gls{qcp}\citep{ruegg_quantum_2008,jain_higgs_2017,hong_higgs_2017,su_stable_2020}, which suppresses the \goldstone bosons and thus stabilizes the \higgs mode.
A second option is to lower the dimensionality\citep{canali_theory_1992,affleck_longitudinal_1992,schulz_dynamics_1996,essler_quasi-one-dimensional_1997,zhou_amplitude_2021}; in quasi-\oned, enhanced fluctuations weaken the long range magnetic order and also reduce the spectral weight of the \goldstone bosons\citep{zhou_amplitude_2021}. 
In \ce{CuBr2}, not only is the system fundamentally one-dimensional, but is thought also to lie in close proximity to a zero-temperature \gls{qcp}\citep{furukawa_ground-state_2012} separating the spiral phase considered here and a paraelectric Haldane dimer phase.
Both of these mechanisms may thus contribute to stabilizing the \higgs electromagnon in \ce{CuBr2}.
 
\section{Conclusion}
To summarize, we have presented evidence of a novel electromagnon arising from the \higgs mode of the spiral magnetic order in \ce{CuBr2}.
This mode appears alongside the pseudo-Goldstone mode in the \gls{trshg} data as a low-frequency oscillation in the longitudinal component of the electric polarization, which softens on warming close to $T_c$.
Looking forward, we note that the two mechanisms we identified for stabilizing this mode in \ce{CuBr2}---low dimensionality and potential proximity to a \gls{qcp}---are not necessarily unique to this material.
Thus, the \higgs electromagnon presented here may in fact be a common feature of quasi-\oned multiferroics, and its observation could indicate a wealth of new opportunities to explore the \higgs mode of particle physics in novel condensed-matter contexts.
 
\section{Methods}\label{sec:methods}
\Gls{trshg} measurements were carried out using a fast-rotating optical grating setup described previously\citep{fichera_second_2020,harter_high-speed_2015,torchinsky_low_2014}.
\qty{100}{fs} ultrashort pulses from a regeneratively amplified \qty{5}{kHz} \ce{Ti}:Sapphire laser were used to pump an \gls{opa}, producing \qty{1300}{nm} (\cref{fig:fig3}) or \qty{1650}{nm} (\cref{fig:fig4}) pump pulses which were delayed with an optical delay line and focused at normal incidence to a \qty{300}{um}-diameter spot on the sample.
The pump fluence was $\sim \qty{1}{mJ\cdot cm^{-2}}$ for all measurements.
A small portion of the \ce{Ti}:Sapphire output was used for the \gls{shg} probe experiment, the output of which was spectrally filtered with a \qty{400}{nm} bandpass filter, collected by a photomultiplier tube, filtered with a lock-in amplifier, and correlated with the plane of incidence angle using an optical rotary encoder.
To measure the pump-induced change in the \gls{shg} signal, the pump pulses were chopped at a frequency of \qty{2.5}{kHz}, and the lock-in amplifier was set to that frequency so as to measure $I_{\mathrm{pump}+\mathrm{probe}}-I_\mathrm{probe}$.
For the pump-probe \gls{rashg} measurements, the plane of incidence was rotated while the delay stage was moved and the polarizers were controlled automatically using homebuilt polarization rotators described in \citet{morey_automated_2024}.
For the single-angle \gls{trshg} measurements, the plane of incidence was parked at the angle which maximized the static \gls{shg} intensity in the respective polarization channel.
 
\section{Acknowledgments}\label{sec:acknowledgments}
This work was supported by the U.S. Department of Energy, BES DMSE award number DE-FG02-08ER46521.
A.K. was supported by the Gordon and Betty Moore Foundation EPiQS Initiative through Grant No. GBMF8684 at the Massachusetts Institute of Technology.
T.S. was supported by the Department of Energy under grant DE-SC0008739.
 
\bibliography{main-expanded}

\clearpage

\fakesection{sup}
\renewcommand{\thepage}{S\arabic{page}} 
\renewcommand{\thesection}{S\arabic{section}}  
\renewcommand{\thefigure}{S\arabic{figure}} 
\setcounter{figure}{0}
\setcounter{section}{0}
\setcounter{subsection}{0}
\setcounter{page}{1}

\title{Supplementary material for ``\Higgs-mode electromagnon in the spin-spiral multiferroic \ce{CuBr2}''}

\maketitle
 
\section{Electromagnons in \ce{CuBr2}}\label{sup:pimpliess}
The observed co-existence of spiral magnetic order and ferroelectricity is due to the spin-orbit coupling enabled interaction term between the spins $\vec S$ and the electronic polarization $\vec P$~\citep{katsura_dynamical_2007}:
\begin{equation}
    H_{s-P} = \lambda \sum_i \vec P_i \cdot \left(\hat x \times \left(\vec S_i \times \vec S_{i+1}\right) \right)
\end{equation}
The ordered state for $T<T_N$ is a multiferroic with spontaneous polarization
\begin{equation}
\left<\vec P_i\right> = P_0 \hat y
\end{equation}
and spiral spin ordering
\begin{equation}
\left< \vec S_i\right> \equiv \vec S_{0,i} = S_0 \left( \cos (\vec Q \cdot \vec R_i) \hat x + \sin (\vec Q \cdot \vec R_i) \hat y\right),
\end{equation}
where $\vec Q$ is the spin-ordering wavevector and $\vec R_i$ are the spatial coordinates of the Cu atoms.
Let us consider fluctuations about this ordered state and ask which fluctuations are detectable via SHG.
Representing fluctuations in the polarization by $\delta \vec P_i$ and spin by $\delta \vec S_i$, we get the following fluctuation Hamiltonian:
\begin{equation}\label{eq:fluctuationhamiltonian}
    H^f_{s-P} = \lambda \sum_i \delta \vec P_i \cdot \left(\hat x \times \left(\delta \vec S_i \times \vec S_{0,i+1} \right)+ \hat x \times \left( \vec S_{0,i} \times \delta \vec S_{i+1}\right) \right) + \mathcal{O}(\delta \vec P^2, \delta \vec S^2).
\end{equation}
Expanding the spin fluctuations along all directions, we find that they couple only to polarization fluctuations along $\hat y$ and $\hat z$.
Focusing on zero-momentum polarization fluctuations (since they are sensitive to SHG),
\begin{equation}\label{eq:ftfluctuationhamiltonian}
\begin{aligned}
\frac{H^f_{s-P}}{\lambda S_0} = &i \sin (\vec Q \cdot \vec a) \delta P_z(\vec q=0) \left(\delta S_z(-\vec Q) - \delta S_z (\vec Q) \right)\\
&+\sin (\vec Q \cdot \vec a) \delta P_y(\vec q=0) \left( -\delta S_x(-\vec Q) + i\delta S_y(-\vec Q) - \delta S_x(\vec Q) - i\delta S_y(\vec Q) \right)\\
& + \mathcal{O}(\delta \vec P^2, \delta \vec S^2)
\end{aligned}
\end{equation}
where $\vec a$ is the lattice vector along the chain.
Transverse polarization fluctuations $\delta P \sim \hat z$ couple to a uniform rotation of the spin-plane about the $x$ axis.
These are the electromagnons discussed in \citet{katsura_dynamical_2007}.
The longitudinal fluctuations, on the other hand, couple to longitudinal fluctuations of the magnetization on each site.
 
\section{Energy of the pseudo-Goldstone mode}\label{sup:anisotropyenergy}
Easy-plane anisotropy in the spin Hamiltonian that penalizes any spin-canting out of the Cu-Br plane imparts a finite energy to the pseudo-\goldstone mode.
In this section, we calculate this energy within \gls{lswt}.
Consider the spin Hamiltonian:
\begin{align}
    H = J_1 \sum_{i} \vec{S}_i \cdot \vec{S}_{i+1} + J_2 \sum_{i} \vec{S}_i \cdot \vec{S}_{i+2} + (J^z_1-J_1) \sum_{i} S^z_i  S^z_{i+1} + (J^z_2 - J_2) \sum_{i} S^z_i S^z_{i+2}
\end{align}
where we have introduced anisotropy terms corresponding to distinct spin couplings along $\hat{z}$.
Let us consider a classical ground state where the spins form a spin spiral state in the $x-y$ plane with an ordering wave-vector $\vec Q$ set by minimizing the classical ground state energy (per Cu atom): $-2J_1 \cos \left(\vec Q \cdot \vec a \right) - 2J_2 \cos \left(2\vec Q \cdot \vec a\right)$.
This gives $\vec Q \cdot \vec a = \cos^{-1} \left(-\frac{J_1}{4J_2}\right)$.
Our next step is to study spin waves about this classical ground state.
It will be convenient to introduce rotated spin variables for a spin located at $\vec R_i$
\begin{align}
    \begin{bmatrix} \tilde{S}^x_i \\ \tilde{S}^y_i \end{bmatrix} = 
\begin{bmatrix} 
\cos \vec{Q} \cdot \vec R_i & \sin\vec{Q} \cdot \vec R_i \\
-\sin\vec{Q} \cdot \vec R_i & \cos\vec{Q} \cdot \vec R_i
\end{bmatrix} 
\begin{bmatrix} S^x_i \\ S^y_i \end{bmatrix}
\end{align}
and $\tilde{S}^z_i = S^z_i$ such that the spin spiral is represented as a uniform ferromagnet along $\hat{y}$, in terms of $\vec{\tilde{S}}$.
The spin Hamiltonian can be rewritten as
\begin{align}
    H =& J_1^z \sum_i \tilde{S}^z_i \tilde{S}^z_{i+1} + J_2^z \sum_i \tilde{S}^z_i \tilde{S}^z_{i+2} \nonumber \\
    & + J_1 \sum_i \tilde{S}^x_i \tilde{S}^x_{i+1} \cos \vec{Q} \cdot \vec a + \tilde{S}^x_i \tilde{S}^y_{i+1} \sin \vec{Q} \cdot \vec a - \tilde{S}^y_i \tilde{S}^x_{i+1} \sin \vec{Q} \cdot \vec a + \tilde{S}^y_i \tilde{S}^y_{i+1} \cos \vec{Q} \cdot \vec a \nonumber \\
    & + J_2 \sum_i \tilde{S}^x_i \tilde{S}^x_{i+2} \cos 2\vec{Q} \cdot \vec a + \tilde{S}^x_i \tilde{S}^y_{i+2} \sin 2\vec{Q} \cdot \vec a - \tilde{S}^y_i \tilde{S}^x_{i+2} \sin 2\vec{Q} \cdot \vec a + \tilde{S}^y_i \tilde{S}^y_{i+2} \cos 2\vec{Q} \cdot \vec a
\end{align}
Now we introduce Holstein-Primakoff bosons $b$ for spin S:
\begin{align}
    \tilde{S}^z_i + i\tilde{S}^x_i &= \sqrt{2S} \left( 1 - \frac{b^{\dagger}_i b_i}{2S}\right)^{1/2} b_i \nonumber \\
    \tilde{S}^y_i &= S - b^{\dagger}_i b_i
\end{align}
In the harmonic approximation, we obtain
\begin{align}
    \frac{H}{S} =& \frac{ J_1^z - J_1 \cos \vec Q \cdot \vec a}{2} b_i b_{i+1} + \frac{ J_2^z - J_2 \cos 2\vec Q \cdot \vec a}{2} b_i b_{i+2} \nonumber \\
    &+\frac{ J_1^z + J_1 \cos \vec Q \cdot \vec a}{2} b^{\dagger}_i b_{i+1} + \frac{ J_2^z + J_2 \cos 2\vec Q \cdot \vec a}{2} b^{\dagger}_i b_{i+2} \nonumber \\
    &-\left(J_1 \cos \vec{Q} \cdot \vec{a} + J_2 \cos 2\vec{Q} \cdot \vec{a} \right) b^{\dagger}_i b_i + \text{h.c.}
\end{align}
Defining 
\begin{align}
A_k =& \cos (\vec{k} \cdot \vec{a}) ( J_1^z + J_1 \cos \vec Q \cdot \vec a )/2 + \cos (2\vec{k} \cdot \vec{a}) ( J_2^z + J_2 \cos 2 \vec Q \cdot \vec a )/2  \nonumber \\
&-J_1 \cos \vec{Q} \cdot \vec a - J_2 \cos 2\vec{Q} \cdot \vec a \nonumber \\
B_k =& \cos (\vec{k} \cdot \vec{a}) ( J_1^z - J_1 \cos \vec Q \cdot \vec a )/2 + \cos (2\vec{k} \cdot \vec{a}) ( J_2^z - J_2 \cos 2 \vec Q \cdot \vec a )/2
\end{align}
we obtain the spin-wave spectrum to be $\omega_k = 2S \sqrt{A_k^2 - B_k^2}$.
In the absence of the easy-plane anisotropy, $\omega_{k=Q} = 0$, and there is a true \goldstone mode.
However, in the presence of the anisotropy, 
\begin{align}
    \omega_{k=Q} =\frac{S}{8} \sqrt{\frac{((J_1 + 4 J_2)^2 (J_1^2 - 4 J_1 J_2 + 8 J_2^2) (-2 J_1 J^z_{1} J_2 + 
   8 J_2^2 (J_2 - J^z_{2}) + J_1^2 (J_2 + J^z_{2})))}{2 J_2^5}}
\end{align}
Using $J_1=J_1^z=\qty{8.8}{meV}$\citep{lebernegg_magnetism_2013}, $J_2=\qty{-22.2}{meV}$\citep{lebernegg_magnetism_2013} and anisotropy $J_2-J_2^z = \qty{0.15}{meV}$\citep{lee_investigation_2012}, we obtain the pseudo-\goldstone mode energy to be \qty{2.54}{meV}. 
Note that deviations from the \gls{lswt} value are expected in the real system due to quantum fluctuations which are not treated here.

Additionally, a zero-energy phason mode exists at $\vec{k}=0$, regardless of the values of the exchange constants. 
 
\section{Fits of time domain signals}\label{sup:timedomain}
\begin{figure}
\phantomsubfloat{\label{fig:timedomaina}}
\phantomsubfloat{\label{fig:timedomainb}}
\centering
\makebox[\linewidth]{\includegraphics[width=183mm]{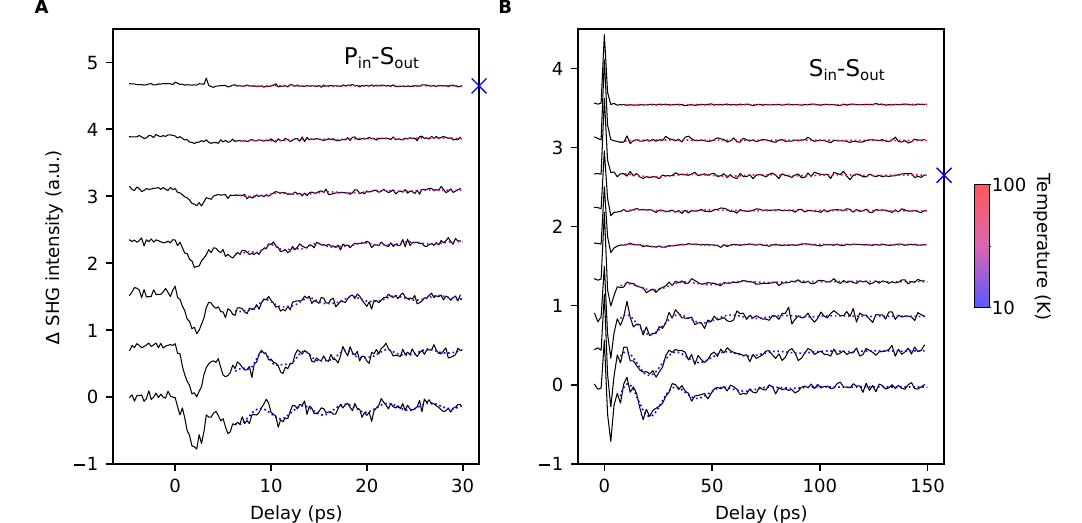}}
\captionsetup{singlelinecheck=off}
\caption[]{
\label{fig:timedomain}
\begin{enumerate*}[label=\caplabel, ref=\capref]
\item[] Time-domain signals corresponding to \item \cref{fig:fig4a} and \item \cref{fig:fig4b}.
Dashed lines depict least-squares fits to the data in a damped harmonic oscillator model, see \supcref{sup:timedomain}.
The blue cross in each figure indicates the transition temperature $T_c \approx \qty{\Tc}{K}$.
\end{enumerate*}
}
\end{figure}

Time-domain plots corresponding to the frequencies in \cref{fig:fig4} are illustrated in \cref{fig:timedomain}.
Each plot is a least-squares fit of the data to a damped harmonic oscillator model
\begin{equation}\label{eq:P0deltaP}
I^\mathrm{SHG}_p(t, \params) = P^0_p\delta P_p(t, \params)+[\delta P_p(t)]^2,
\end{equation}
where
\begin{equation}\label{eq:model}
\delta P_p = A_pe^{-\gamma_p t}\cos\left(\sqrt{(2\pi\nu_p)^2-\gamma_p^2}t+\psi_p\right),
\end{equation}
$p \in \{\mathPS, \mathSS\}$, and $\params$ denotes the set of free parameters to be estimated.

\begin{figure}
\centering
\makebox[\linewidth]{\includegraphics[width=89mm]{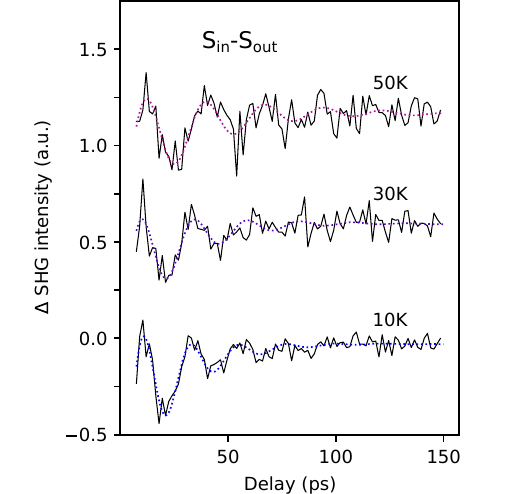}}
\captionsetup{singlelinecheck=off}
\caption[]{
\label{fig:timedomain_zoom}
\begin{enumerate*}[label=\caplabel, ref=\capref]
\item[] Rescaled \SS time-domain signals (see \cref{fig:timedomainb}) for select temperatures below $T_c$.
\end{enumerate*}
}
\end{figure}

\begin{figure}
\centering
\makebox[\linewidth]{\includegraphics[width=183mm]{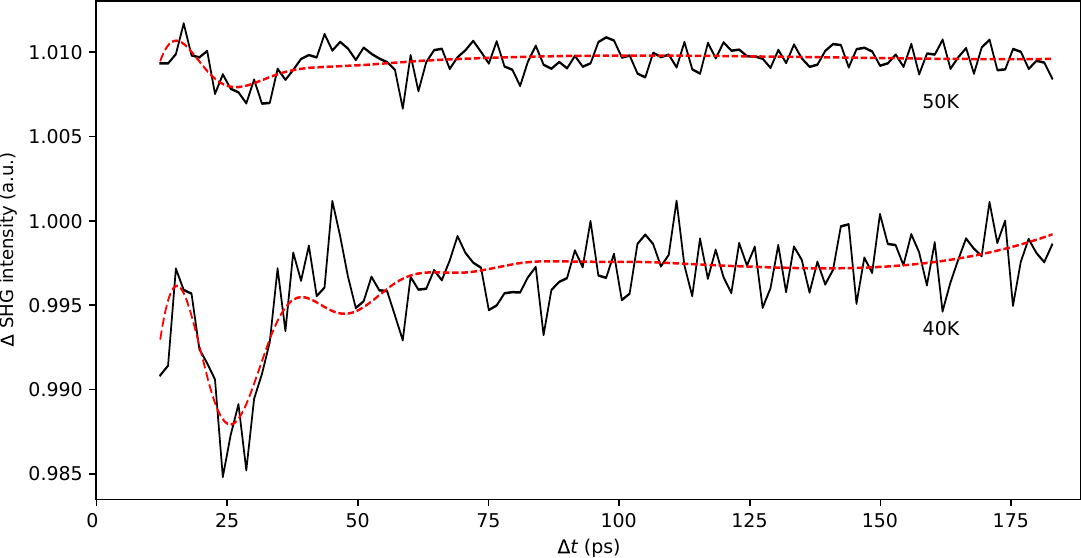}}
\captionsetup{singlelinecheck=off}
\caption[]{
\label{fig:timedomain_nosoften}
\begin{enumerate*}[label=\caplabel, ref=\capref]
\item[] \SS time-domain signals (see \cref{fig:timedomainb}) for select temperatures approaching $T_c$.
Dashed lines depict least-squares fits to the data in a variant of \cref{eq:model} where $\nu_\mathrm{SS}$ is constrained not to vary with temperature.
\end{enumerate*}
}
\end{figure}

The main conclusion of these fits is that the frequency of the two modes (most notably, the low-frequency \SS mode) soften on approaching $T_c$.
This may also be seen heuristically from the time-domain signals without doing any fits.
\Cref{fig:timedomain_zoom} shows an enlarged (i.e., scaled to account for the decrease in signal amplitude) view of the \SS signal for three temperatures below $T_c$, showing a clear decrease in the oscillation frequency at high temperature.
\Cref{fig:timedomain_nosoften} shows an alternative fit where the frequency parameter $\nu_\mathrm{SS}$ is constrained to be constant as a function of temperature, showing that our data is not consistent with a hypothetical model where the frequency shift with temperature in \cref{eq:model} is only attributed to the damping term $\gamma_\mathrm{SS}$.

\section{Error bars in \cref{fig:fig4}}\label{sup:errorbars}
In this section, we describe how the uncertainties in the least square estimates of the frequency parameter $\nu$ of \cref{eq:model}, which are depicted as a function of temperature in \cref{fig:fig4}, were calculated from the the time-domain signals in \cref{fig:timedomain}.
For each temperature and polarization channel, a \gls{lm} algorithm was used to find the minimum $\params_0$ of the objective function
\begin{equation}\label{eq:objectivefunction}
f_p(\theta) \propto \sum_{n=0}^{N-1}\left(I^\mathrm{SHG}_p(t_n, \params) - I^\mathrm{SHG}_{p,n}\right)^2,
\end{equation}
where $\{(t_n, I^\mathrm{SHG}_{p,n}), n\in (0, 1, \ldots, N-1)\}$ are the data points in \cref{fig:timedomain}, and we have assumed the noise level is independent of delay.
The uncertainty in each parameter is estimated within a parametric bootstrap\citep{dekking}: for each temperature, \num{1000} bootstrap samples are generated by adding noise (normally distributed, with variance given by the variance of data points at long times where the signal is constant) to the \gls{lm} estimate $I^\mathrm{SHG}_p(t_n, \params_0)$.
For each bootstrap sample $s$, an estimate $\params_s$ is computed by minimizing \cref{eq:objectivefunction} as above, and the \qty{95}{\percent} confidence interval reported in \cref{fig:fig4} is taken to be \num{1.96} times the standard deviation of the distribution $\{\params_s-\theta_0\}$.
 
\section{Fits to static RASHG data}\label{sup:static}
\begin{figure}
\centering
\makebox[\linewidth]{\includegraphics[width=89mm]{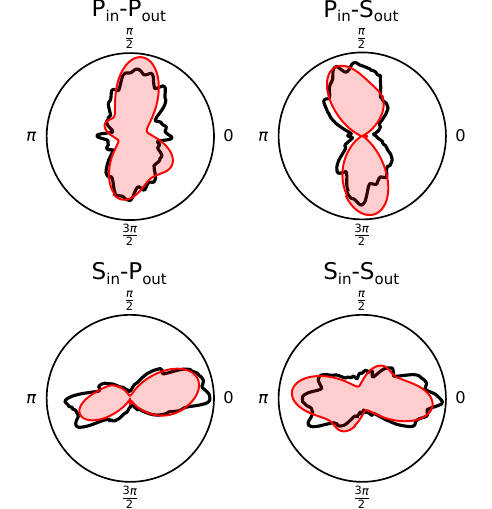}}
\captionsetup{singlelinecheck=off}
\caption[]{
\label{fig:static}
\begin{enumerate*}[label=\caplabel, ref=\capref]
\item[] Fits (red) to static \gls{rashg} data (black) depicted in \cref{fig:fig3}.
\end{enumerate*}
}
\end{figure}

The static \gls{shg} intensity was fit by \cref{eq:shgintensityequation}.
The susceptibility tensor was taken to be the form \cref{eq:susceptibilityshortform}, plus an additional $C_1$ component (likely due to surface adsorbates).
The result is shown in \cref{fig:static}.

\section{Disappearance of oscillations above $T_c$}\label{sup:nooscillationabovetc}
In the main text, we noted that the fact that both modes disappear above $T_c$ is consistent with the fact that the macroscopic polarization $\vec{P}_0$ also disappears above this temperature. 
To see this, note again that the oscillation in the \gls{trshg} signal is amplified by the static polarization component $P_0$ via \cref{eq:P0deltaP}, which makes these modes more apparent at low temperatures.
Despite this, we do expect fluctuations about $P_0=0$ in the material to exist above $T_c$; they are simply not observable in our experiment.
 
\section{Excitation mechanism\label{sup:excitationmechanism}}
While the exact mechanism responsible for the ultrafast excitation of electromagnons shown in this work cannot be unambiguously identified with our data, we describe one possible explanation here.
Ultimately the coupling between the polarization and spin degrees of freedom is due to the spin-orbit interaction; that is, fluctuations in the polarization are essentially due to changes in the relative occupancy of different local electron orbitals which couple to the electron spin via \cref{eq:scrosss}.
We may thus imagine a scenario where the light pulse interacts with the material via the orbital degrees of freedom (for example via \gls{isrs}, which is allowed to excite odd-parity modes in \ce{CuBr2} since inversion symmetry is broken below the multiferroic transition).
These orbital fluctuations then, in turn, drive electromagnons through the coupling between polarization and spin fluctuations described here.
 
\section{Excluded possibilities for observed results}\label{sup:excluded}
\subsection{$\delta \vec{P} || \hat{x}$ oscillation}\label{sup:nopxmode}

Without loss of generality, let the maximum of the SHG in \SP occur when the incoming electric field is along $\hat{x}$.
Then, we have:
\begin{equation}
\Delta I_\mathrm{SP}^\mathrm{SHG} \propto |\hat{e}^\mathrm{out}_i \chi_{ijkl}\hat{e}^\mathrm{in}_j\hat{e}^\mathrm{in}_k[P_{0l}+\delta P_l]|^2-|\hat{e}^\mathrm{out}_i \chi_{ijkl}\hat{e}^\mathrm{in}_j\hat{e}^\mathrm{in}_kP_{0l}|^2
\end{equation}
with $\hat{e}^\mathrm{in}_i || x$ and $P_{0l} || y$, we have
\begin{equation}
\label{eq:longdeltai}
\Delta I_\mathrm{SHG} \propto 2\hat{e}^\mathrm{out}_i \hat{e}^\mathrm{out}_j\chi_{ixxy} \chi_{jxxx}P_{0y}\delta P_x + \hat{e}^\mathrm{out}_i \hat{e}^\mathrm{out}_j\chi_{ixxx} \chi_{jxxx}\delta P_x\delta P_x 
\end{equation}
Since we are in \SP, $\hat{e}^\mathrm{out}_i \perp x$; thus, \cref{eq:longdeltai} involves the tensor elements $\chi_{yxxx}$ and $\chi_{zxxx}$.
Both of these elements are zero due to the $x \rightarrow -x$ mirror symmetry.
Thus, the $\delta \vec{P} || \hat{x}$ oscillation is not visible in our experiment.

\subsection{Coherent acoustic phonons}
Another possible source of coherent oscillations in the SHG signal are acoustic phonons.
We first consider modes with wavevectors determined by the wavelength of the probe pulse, which are close to the zone center.

\subsubsection{Near the zone center}
\label{sup:gamma_phonons}
Acoustic phonon modes near the zone center are known to generate coherent oscillations in some materials \citep{thomsen_surface_1986,wu_femtosecond_2007,ezzahri_coherent_2007,hortensius_coherent_2021}.
If the pump pulse excites these modes, their presence imposes a Bragg condition on the probe pulse \citep{zhao_detection_2011}.
The corresponding oscillation frequency is given by
\begin{align}
    \omega_{\text{ph}} = 2c_s n \cos \theta'/\lambda
\end{align}
where $\theta'$ is the refracted angle of incidence: $\sin \theta' = (1/n) \sin \theta$.
Here, $c_s$ is the speed of sound, $n$ is the refractive index and $\lambda$ is the wavelength of the probe light pulse.
Using $\lambda = \qty{1300}{nm}$, $n=\num{2.1}$, $\theta = \ang{10}$, and from inelastic neutron scattering data \citep{wang_observation_2017}, $c_s \approx \qty{0.975}{meV\cdot nm}$, we obtain $\omega_{\text{ph}} = \qty[exponent-mode=scientific]{0.00075}{THz}$.
From \cref{fig:fig4b}, the experimental value at low temperatures is $\approx \qty{0.05}{THz}$, which is two orders of mangnitude larger, thus ruling out zone-center acoustic phonons as the origin of the observed oscillations.

\begin{figure}
\centering
\makebox[\linewidth]{\includegraphics[width=183mm]{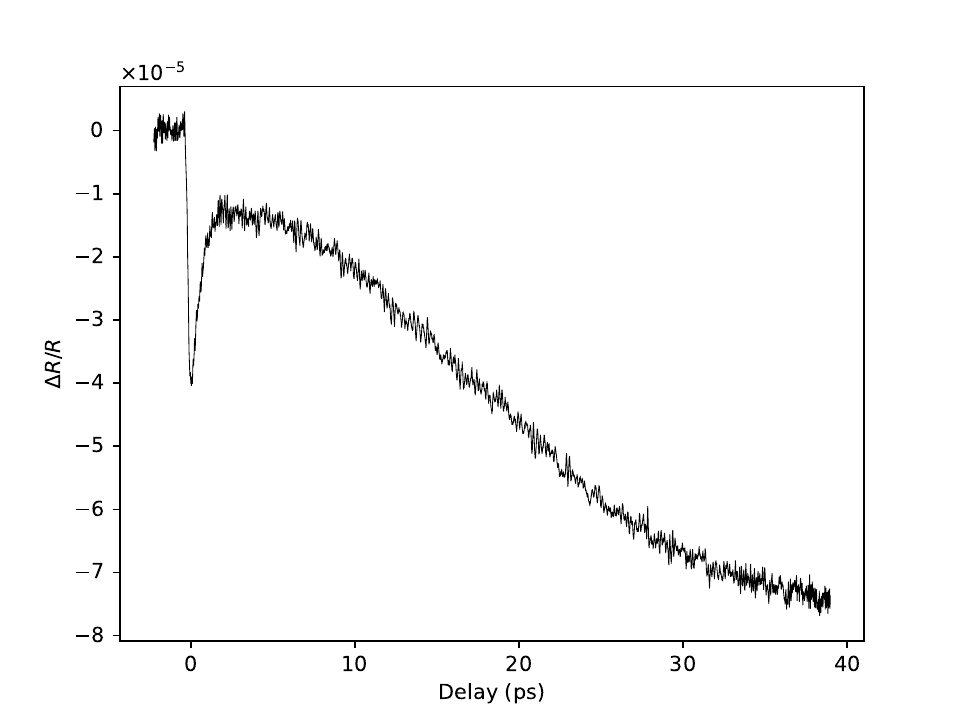}}
\captionsetup{singlelinecheck=off}
\caption[]{
\label{fig:pumpprobe}
\begin{enumerate*}[label=\caplabel, ref=\capref]
\item[] Pump-probe reflectivity at \qty{30}{K} in \ce{CuBr2} showing no discernible coherent oscillations over a \qty{40}{ps} interval.
\end{enumerate*}
}
\end{figure}

Furthermore, coherent phonons excited near the zone center should be observed in time-resolved reflectivity, which is not true of the modes considered in this work (see \cref{fig:pumpprobe}).
Note that transient reflectivity measurements involve (in our case) an \qty{80}{MHz} repetition rate and are thus orders of magnitude more sensitive to small changes in optical constants than time-resolved \gls{shg} \citep{hsieh_selective_2011}.
The fact that the modes observed in this work appear in \gls{shg} and not in reflectivity thus suggests that they primarily modulate the electric polarization itself rather than the surrounding lattice, which rules out coherent phonons.

\subsubsection{Zone-folded acoustic phonons}\label{sup:phonons}

\begin{figure}
\centering
\makebox[\linewidth]{\includegraphics[width=183mm]{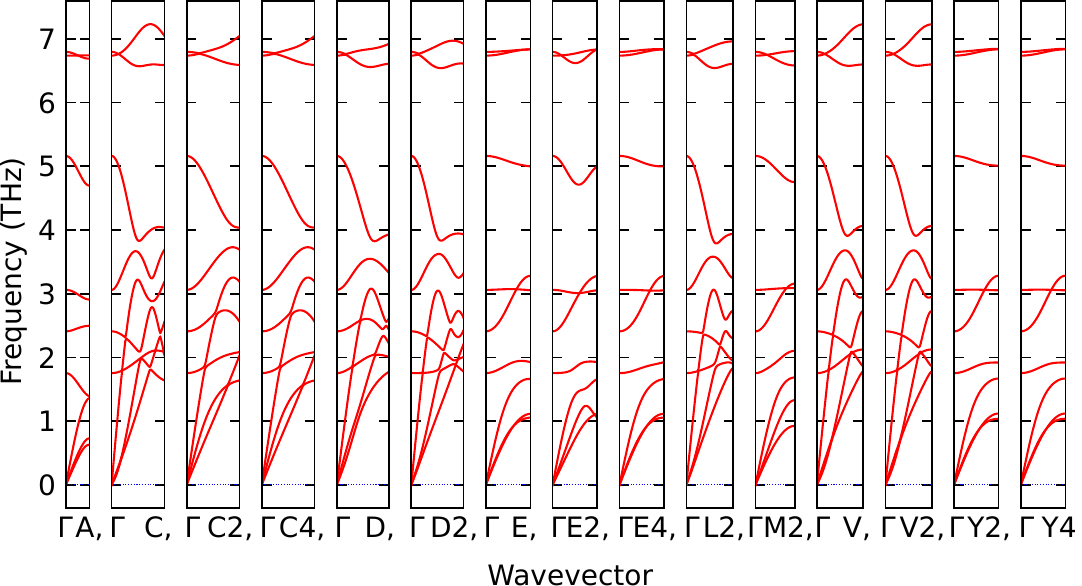}}
\captionsetup{singlelinecheck=off}
\caption[]{
\label{fig:phonons}
\begin{enumerate*}[label=\caplabel, ref=\capref]
\item[] Phonon band structure of \ce{CuBr2} within a finite displacement calculation.
\end{enumerate*}
}
\end{figure}

Phonon band structure calculations were carried out using the finite displacement method\citep{togo_first_2015} with a distance of \qty{0.01}{\angstrom} within a $3\times3\times3$ supercell.
Forces were calculated via the DFT-D2 method\citep{grimme_semiempirical_2006} and LDA+U method\citep{dudarev_electron-energy-loss_1998} ($U_\mathrm{\ce{Cu}}=\qty{3}{eV}$) using a $7\times7\times5$ $k$-mesh with \num{122} irreducible $k$-points and a plane-wave cutoff energy of \qty{100}{eV}.
The result is shown in \cref{fig:phonons}.
The acoustic phonons in \cref{fig:phonons} all disperse too rapidly for the \qty{0.05}{THz} oscillation in the \gls{trshg} to be consistent with a zone-folded (at $k=(0, 0.235, 0.5)$) acoustic phonon (which may occur due to, for example, some static coupling between the lattice and spin degrees of freedom).

\subsection{Magnetic SHG}\label{sup:magshg}

In principle, magnetic systems with broken inversion symmetry may generate electric-dipole SHG with or without a static electric dipole moment.
In this section, we wish to show that this is not the case in \ce{CuBr2}; i.e., in \ce{CuBr2}, the SHG intensity is a direct measure of the macroscopic electric dipole moment.

Indeed, in the presence of such a static electric dipole moment $\vec{P}_0$, we typically expect the SHG response to be directly proportional to it; i.e.
\begin{equation}\label{eq:shgintensityequation}
I(2\omega) \propto |\hat{e}^\mathrm{out}_i \chi_{ijk} \hat{e}^\mathrm{in}_j \hat{e}^\mathrm{in}_k|^2,
\end{equation}
where
\begin{equation}\label{eq:susceptibilityshortform}
\chi_{ijk} = \chi_{ijkl} P_{0l},
\end{equation}
and $\hat{e}^\mathrm{in}$, $\hat{e}^\mathrm{out}$ are unit vectors in the direction of the incoming and outgoing electric fields, respectively.
In \ce{CuBr2}, we have  
\begin{equation}
\vec{P}_0 = \sum\limits_{\left<i, j\right>} \hat{x} \times \left(\vec{S}_i \times \vec{S}_j\right),
\end{equation}
i.e., the static polarization is quadratic in the spin degree of freedom.

The question, then, is whether there exists some additional term
\begin{equation}
\chi'_{ijk} = \chi_{ijkl} G_{0l},
\end{equation}
where $\vec{G}_0$ is either (a) linear in spin, or (b) quadratic in the spins but not of the form $\sum_{\left<ij\right>}\vec{S}_i\times\vec{S}_j$.
For case (b), note that the term $\sum_{\left<ij\right>}\vec{S}_i\times\vec{S}_j$ is the only quadratic form which is simultaneously antisymmetric in the bond direction and $\vec{q}=0$ (i.e. each bond has the same coefficient).

For case (a), we argue here that any such term is weak due to the approximate time-reversal symmetry of the spiral magnetic order.
Consider first a four-site commensurate approximant of the incommensurate spin spiral.
This phase has a symmetry element consisting of the time-reversal operation followed by a translation by half of the magnetic supercell.
Thus, the point group contains time-reversal symmetry.
Since $\vec{G}_0$ is linear in spin, time-reversal takes $\vec{G}_0\rightarrow-\vec{G}_0$; but since time-reversal is a symmetry, it must also take $\chi'_{ijk}\rightarrow\chi'_{ijk}$ and $\chi_{ijkl}\rightarrow\chi_{ijkl}$.
Thus, $\chi'_{ijk}=0$ in the commensurate approximation.

In the incommensurate case, note that the magnetic point group of an incommensurate magnetic phase is defined as the set of point-group operations present in the operations belonging to the superspace group.
Thus, for a single-k incommensurate magnetic structure, time-reversal is always an element of the magnetic point group.
This is due to the fact that the lattice constant in the chain direction is $3.51$ \si{\angstrom}, so lengthscales associated with translations in the space group are much smaller than the probe wavelength ($\sim\qty{800}{nm}$).
The symmetry group ``seen'' by the probe thus contains time reversal to a very good approximation.

\subsection{Multi-phason excitation}

While the amplitude mode of the spin spiral in \ce{CuBr2} is the only single-particle excitation that couples to $\delta P_y$ (see \cref{eq:ftfluctuationhamiltonian}), multiparticle excitations are, in principle, also allowed.
Our system features gapless phason modes corresponding to spin rotations about $\hat{z}$, where the zero-momentum mode represents a global shift of the spin angle within the $x$-$y$ plane, incurring zero energy cost.
The relevant energy scale that determines the phason sound velocity is set by intra-chain coupling terms, which are on the order of $\sim\qty{10}{meV}$.
Consequently, the peak in the phason joint density of states occurs at this high-energy scale, far exceeding the $\qty{0.2}{meV}$ energy of our observed low-frequency oscillation.
Multi-phason excitations are thus inconsistent with the long-lived oscillation detected in our experiment.
A common decay channel for Higgs mode excitations is into low-energy Goldstone modes \citep{jain_higgs_2017}.
However, we anticipate that decay into phason excitations is suppressed in this case due to the energetics outlined above.

\subsection{Amplitude mode in Canted spin-spirals}

In materials where the spins are canted out of the plane at an angle, resulting in finite magnetization, transverse spin oscillations can induce longitudinal oscillations in the polarization \citep{takahashi_versatile_2016}.
However, neutron scattering experiments indicate that the spins in \ce{CuBr2} lie strictly within a plane \citep{lee_investigation_2012}.
Additionally, these experiments required the application of strong magnetic fields ($\sim 5$ T) to observe the effect \citep{takahashi_versatile_2016}.
Consequently, this mechanism is not expected to play a role in the present context.
 
\end{document}